# Deterministic formation of interface states in some two-dimensional photonic crystals with conical dispersions

Xueqin Huang, Meng Xiao, Zhao-Qing Zhang, C. T. Chan[*]

Department of Physics and Institute for Advanced Study, Hong Kong University of Science and Technology, Clear Water Bay, Hong Kong, China

[*]Corresponding author: phchan@ust.hk

## Abstract

There is no assurance that interface states can be found at the boundary separating two materials. As a strong perturbation typically favors wave localization, it is natural to expect that an interface state should form more easily in the boundary that represents a strong perturbation. Here, we show on the contrary that in some two dimensional photonic crystals (PCs) with a square lattice possessing Dirac-like cone at k=0, a small perturbation guarantees the existence of interface states. More specifically, we find that single-mode localized states exist in a deterministic manner at an interface formed by two PCs each with system parameters slightly perturbed from the conical dispersion condition. The conical dispersion guarantees the existence of gaps in the projected band structure which allows interface states to form and the assured existence of interface states stems from the geometric phases of the bulk bands.

Conical dispersions in periodic systems can give rise to many novel properties in electronic and classical wave systems [1-10]. The most famous example is graphene's Dirac cones at the K and K′ points of the Brillouin zone [1-3]. The existence of these conical dispersions at the zone boundary is guaranteed by space group symmetry. It was shown recently that conical dispersions can also be realized at **k**=0 in 2D photonic and phononic crystals, but instead of group symmetry, the existence of a conical dispersion at the zone center requires accidental degeneracy [8-12]. For example, if we can arrange system parameters so that there is accidental degeneracy of monopole and dipole excitations, we can get conical dispersion at **k**=0 for 2D PCs with a square lattice [11]. Using effective medium theory, such 2D PCs have been shown to possess effective $\varepsilon = \mu = 0$ at the frequency of Dirac-like point. Many phenomena that are specific to zero-refractive-index materials [13-21], such as wave tunneling through arbitrary-shaped waveguides, transforming wavefronts and cloaking inside waveguides, can be achieved by using 2D PCs with conical dispersion at **k**=0 [11].

In this paper, we show that conical dispersion at **k**=0 has yet another important implication: it can give rise to interface states in a deterministic manner. The TM polarization bulk band structure, with electric field along cylinder axis, of a PC with $C_{4v}$ symmetry (shown schematically in the inset of Fig. 1(a)) possessing a conical dispersion at **k**=0 is shown in Fig. 1(a). The dispersion has a upper cone and lower cone meeting at a Dirac point with $f = 0.541 c/a$. Here, $c$ is the speed of light in vacuum, $a$ is the lattice constant. There is an additional band that is flat near **k**=0. The states in this flat band are quasi-longitudinal, with the average **H** field parallel to **k** [11]. The projected band structure with $k_\parallel$ along [01] (or [10]) direction is shown in Fig. 1(b). The conical dispersion guarantees that there are gaps in the projected band structure both above and below the Dirac-like point. These gaps in the projected band are separated by the allowed states coming from the quasi-longitudinal band. We will show that the gaps above and below the quasi-longitudinal bands have "opposite characters" stemming from the geometric properties of the bulk bands. These geometric properties guarantee the existence of interface states in an interface along the [01] direction formed between two PCs with $C_{4v}$ symmetry which have their system parameters (dielectric constant and filling ratio) slightly perturbed from the accidental degeneracy condition required for conical dispersion. Single-mode interface states will always appear in certain gap regions shared by the two PCs. The dispersions of these interface states can be predicted by calculating the surface impedance of each semi-infinite PC. In addition, we show that the surface impedance, as a surface property of a semi-infinite PC, can be obtained from the scattering theory and that the sign of the surface impedance (which determines the existence of the interface states) is determined by the geometric phases of the bulk photonic bands. Our results demonstrate that the surface scattering and the geometric properties of the photonic dispersion of the bulk periodic crystal are related and such knowledge can facilitate the design of localized boundary/interface modes in classical wave systems.

Our system is shown schematically in Fig. 2(a). Two semi-infinite 2D PCs comprising dielectric cylinders in a square lattice with the same lattice constant $a$ are put together to construct an interface along the $y$-direction ([01] direction). The relative permittivities and radii of cylinders PCs on the left half space (red) and right half space (blue) are $\varepsilon_1, R_1$ and $\varepsilon_2, R_2$ respectively. The relative permeability is always $\mu = 1$. The bulk photonic band structures of the two PCs with **E** field along the cylinder axis with $\varepsilon_1 = 10$, $R_1 = 0.205a$, and $\varepsilon_2 = 12.5$, $R_2 = 0.22a$ are shown in Fig. 2(b). These parameters are chosen to be slightly perturbed from the conical dispersion formation condition at **k**=0 [which is $\varepsilon = 12.5$, $R = 0.2a$]. The perturbation will modify the dispersion near the Dirac-like cone, which comprises two linear bands and a quasi-longitudinal band which is nearly flat near **k**=0. It also breaks the triply degenerate states at **k**=0 into a pair of doubly degenerate states and a singlet state. To study the existence of interface states, we calculate the projected band structures of the two PCs along the $y$-direction. The results are shown in Fig. 2(c), where the red color stands for the projected band structure of the PC with $\varepsilon_1 = 10, R_1 = 0.205a$ and the blue color stands for the PC with $\varepsilon_2 = 12.5, R_2 = 0.22a$. There are regions of pass bands, which are marked in red or blue colors for each PC. The regions of band gaps are in white color. There exist a few regions of common gaps for both projected band structures.

To search for the possibility of interface states inside the regions of common gaps, we have calculated the eigen modes of a large slab consisting of the two PCs, each with 15 cylinders along the *x*-direction. Perfectly matched layer boundary conditions are applied to the *x*-direction, whereas a periodic boundary condition is applied to the *y*-direction for each wave vector $k_\parallel$ along the $\overline{\Gamma}\overline{X}$ direction. In two regions of common gaps, we found two branches of interface states near the frequencies $0.53c/a$ and $0.45c/a$. These interface states are marked by green lines in Fig. 2(c). In Fig. 2(d), we plot the **E** field distribution of a typical interface mode in the upper branch marked by a black star on the green line. It is obvious that the interface state is localized near the interface. We found that as long as the structural parameters of PCs on either side of the interface are slightly perturbed relative to conical dispersion formation condition, the existence of interface states is assured independent of the details of the perturbation. (See Appendix for interface state formation for different types of perturbation).

The results of interface states shown above raise an interesting question. Why should a small perturbation near the Dirac-like cone at **k**=0 give rise to interface states and why should such localized states exist deterministically in certain common gaps but absent in others? To answer this question, we first note that the condition for the formation of an interface state is given by $Z_L(\omega, k_\parallel) + Z_R(\omega, k_\parallel) = 0$ [22], where $Z_{L(R)}(\omega, k_\parallel)$ is the surface impedance of the semi-infinite PC on the left (right) for a given $k_\parallel$. The governing parameter is hence the surface impedances of the two semi-infinite PCs inside their common gaps. In order to obtain the surface impedance, we use scattering theory [22-23] and treat the 2D PC as a large stack of one-dimensional (1D) PCs (each 1D layer has the configuration shown in Fig. A5 of Appendix). The impedance parameter, $Z_N(\omega, k_\parallel)$ of a stack of N layers can be obtained from the calculated transmission and reflection coefficients for a N-layer thick slab. It is interesting to point out that if we consider only zero-order inter-layer diffraction, namely that the evanescent wave coupling between adjacent layer can be ignored, the function $Z_N(\omega, k_\parallel)$ is independent of the layer number N [see Appendix]. The surface impedance of a semi-infinite PC $Z(\omega, k_\parallel)$ can hence be obtained from the scattering theory of just one bulk layer (shown in Fig. A5 of Appendix). If we consider only the monopole and dipole bands, the reflection and transmission coefficients of a one-layer PC illuminated by an external plane wave with a given $k_\parallel$ can be obtained analytically and the results are: $\quad r = \dfrac{i}{2ak_x}\left(\tilde{P} + \tilde{M}_x k_\parallel + \tilde{M}_y k_x\right) \quad$ and $\quad t = \left[1 + \dfrac{i}{2ak_x}\left(\tilde{P} + \tilde{M}_x k_\parallel - \tilde{M}_y k_x\right)\right] \quad$ with

$$\tilde{P} = \dfrac{1/\alpha_M + F_6 - ik_\parallel F_3}{\left[\varepsilon_0/(\alpha_E k_0^2) - F_1\right]\left(1/\alpha_M + F_6\right) + F_3^2} \;,\quad \tilde{M}_x = \dfrac{-iF_3 + k_\parallel\left[\varepsilon_0/(\alpha_E k_0^2) - F_1\right]}{\left[\varepsilon_0/(\alpha_E k_0^2) - F_1\right]\left(1/\alpha_M + F_6\right) + F_3^2} \;,\text{ and }\tilde{M}_y = -\dfrac{k_x}{1/\alpha_M + F_4}.$$

Here, $\alpha_E$ and $\alpha_M$ are the monopole and dipolar polarizability of the cylinder, $k_\parallel$ is the wave vector along y-direction, $k_x$ is the wave vector perpendicular to the interface, $k_0 = \omega/c$ where $\omega$ is the angular frequency, and $F_1, F_3, F_4, F_6$ are 1D lattice sums defined as:

$$F_1 = \sum_{m\neq 0}\dfrac{i}{4}H_0\left(k_0|\vec{r} - ma\hat{y}|\right)e^{imk_\parallel a}\bigg|_{\vec{r}=0} \;,\quad F_3 = \dfrac{\partial}{\partial y}\sum_{m\neq 0}\dfrac{i}{4}H_0\left(k_0|\vec{r} - ma\hat{y}|\right)e^{imk_\parallel a}\bigg|_{\vec{r}=0} \;,$$

$$F_4 = \dfrac{\partial^2}{\partial^2 x}\sum_{m\neq 0}\dfrac{i}{4}H_0\left(k_0|\vec{r} - ma\hat{y}|\right)e^{imk_\parallel a}\bigg|_{\vec{r}=0} \;,\quad F_6 = \dfrac{\partial^2}{\partial^2 y}\sum_{m\neq 0}\dfrac{i}{4}H_0\left(k_0|\vec{r} - ma\hat{y}|\right)e^{imk_\parallel a}\bigg|_{\vec{r}=0} \;,\quad \text{where } m \text{ is an}$$

integer, and $H_0(x)$ is the zero-order Hankel function of the first kind. The calculated $r$ and $t$

allow us to define an impedance $Z_1(\omega, k_\parallel) = \pm \dfrac{\sqrt{(r+1)^2 - t^2}}{\sqrt{1 - k_\parallel^2/k_0^2}\sqrt{(r-1)^2 - t^2}}$ for a single layer. The sign of $Z_1(\omega, k_\parallel)$ can be determined by the causality considerations. To see whether the zero-order inter-layer diffraction is an adequate approximation, we compare the projected band structures obtained by full-wave calculation with the multiple scattering calculation with only zero-order inter-layer diffraction (see Fig. A6 of Appendix). The results agree well with each other in our interested frequency region near the Dirac-like cone frequency. The projected band structures of two PCs calculated by scattering theory with only zero-order inter-layer diffraction are shown in Figs. 3(a) and 3(b). Regions in which higher order diffractions play an important role are shaded in purple in Figs. 3(a) and 3(b). We emphasize that the surface impedance parameter is uniquely defined as long as the zero-order inter-layer scattering approximation is valid. In lossless materials, $\text{Im}(Z(\omega, k_\parallel)) = 0$ in the pass band, while inside a gap $Z(\omega, k_\parallel)$ is pure imaginary, i.e., $\text{Im}(Z(\omega, k_\parallel)) > 0$ or $\text{Im}(Z(\omega, k_\parallel)) < 0$ as electromagnetic waves exponentially decay inside the PC. Figures 3(a) and 3(b) show that different gaps near the Dirac-like cone frequency carry different signs of $\text{Im}(Z(\omega, k_\parallel))$ and the sign can be used to label the "characters" of the gap. In particular, we note that for a given $k_\parallel$, the gaps above and below the quasi-longitudinal band always have a different sign of $\text{Im}(Z(\omega, k_\parallel))$. For a given $k_\parallel$, the value of $\text{Im}(Z(\omega, k_\parallel))$ decreases monotonically from 0 to $-\infty$ with increasing frequency in a region with $\text{Im}(Z(\omega, k_\parallel)) < 0$, whereas the value of $\text{Im}(Z(\omega, k_\parallel))$ decreases monotonically from $+\infty$ to 0 with increasing frequency in a region with $\text{Im}(Z(\omega, k_\parallel)) > 0$. This property, together with interface state formation condition of , $\text{Im}(Z_L) + \text{Im}(Z_R) = 0$ implies that there must exist one and only one interface state inside the common gap if the surface impedances of the two PCs have different signs and there cannot be interface states inside the gap if $\text{Im}(Z_L)$ and $\text{Im}(Z_R)$ have the same sign. In Figs. 3(a) and 3(b), we mark explicitly the sign of $\text{Im}(Z(\omega,k_\parallel))$ in each gap of the two PCs. According to the condition of interface state formation, it is easy to see that there are interface states in the common gaps with frequencies around $0.53 c/a$ and $0.45 c/a$, whereas no interface states are allowed in the other common gaps. In addition, we can use the value of $\text{Im}(Z(\omega,k_\parallel))$ extracted from the scattering theory to calculate the interface wave dispersions. The results are shown as black circles in Fig. 3(c). For comparison, we have also carried out the full-wave calculations to obtain the band dispersions of the interface states. These results are shown by green lines in Fig. 3(c). Excellent agreements have been found between the two calculations.

The guaranteed existence of surface or interface states is frequently related to the topological properties of the bulk bands [24-35]. To give an "geometric" interpretation of the formation of interface states, we first note that the projected band structure for a particular $k_\parallel$ comes from the bulk bands with a fixed $k_y = k_\parallel$ and with $k_x$ varying from $-\pi/a$ to $\pi/a$. For example, the pass bands and forbidden gaps in the projected band structure at $k_\parallel = 0.6\pi/a$ in Fig. 3(c) (marked by yellow dashed line) correspond to a reduced 1D band structures shown in Fig. 4 where we plot the reduced 1D band structures of two PCs along $k_x$ direction with $k_y = 0.6\pi/a$ and $\varepsilon_1 = 10, R_1 = 0.205a$, and $\varepsilon_2 = 12.5, R_2 = 0.22a$, respectively.

We calculate the Zak phase [36] of the reduced 1D bands shown in Fig. 4 using the formula

$$\varphi_n = i \int_{-\pi/a}^{\pi/a} \left\langle u_{nk_x,k_y=k_\parallel} \middle| \varepsilon(\vec{r})\partial_{k_x} \middle| u_{nk_x,k_y=k_\parallel} \right\rangle dk_x \ ,$$

where $u_{n\vec{k}}$ is the cell periodic part of the Bloch function of the **E** field for the $n^{th}$ band at a particular $\vec{k}$, $\varepsilon(\vec{r})$ is position dependent relative permittivity. The Zak phase is calculated using the periodic gauge and the origin is chosen at the left boundary of the unit cell as shown in the right panel in Fig. 4(a). The Zak phases for the four lowest bands are $\varphi_1 = \pi$ and $\varphi_n = 0$ for $n=2, 3, 4$. In fact, the zero Zak phase of bands 2 and 3 is required by the $C_{4v}$ symmetry of the PCs (see appendix). Now we apply a rigorous relation found in 1D system with inversion symmetry [37] here. The relation relates the ratio of the signs of $\text{Im}(Z(\omega,k_\parallel))$ in two adjacent gaps, say the $n^{th}$ and $(n-1)^{th}$ gaps, to the Zak phase of the band in between [37], i.e.,

$$\frac{Sgn\left[\text{Im}\left[Z_n(\omega,k_\parallel)\right]\right]}{Sgn\left[\text{Im}\left[Z_{n-1}(\omega,k_\parallel)\right]\right]} = e^{i(\varphi_{n-1}+\pi)} \ . \tag{1}$$

It is easy to show that the sign of $\text{Im}(Z(\omega,k_\parallel))$ in the lowest gap is always negative. The knowledge of the Zak phases of the bulk bands allow us to determine the signs of the gaps through Eq. (1). The results are shown in Figs. 4(b) and 4(c), where the blue color denotes the gaps with $\text{Im}(Z(\omega,k_\parallel)) < 0$ and the red color stands for the gaps with $\text{Im}(Z(\omega,k_\parallel)) > 0$. These results are consistent with that of the scattering theory shown in Figs. 3(a) and 3(b) along the yellow dashed lines, respectively. Thus, by knowing the bulk Zak phase we can also determine the sign of $\text{Im}(Z(\omega,k_\parallel))$ without doing the cumbersome calculation using scattering theory. From Figs. 4(b) and 4(c), it is also seen that there are actually two overlapping gaps that have different signs of $\text{Im}(Z(\omega,k_\parallel))$. The overlap of the second gap in Fig. 4(b) and the third gap in Fig. 4(c) gives rise to the lower branch of the interface states near the frequency $0.45c/a$ found in Fig. 3(c) and the overlap of the third gap in Fig. 4(b) and the fourth gap in Fig. 4(c) gives rise to the higher branch near the frequency $0.53c/a$.

In summary, we show that interface states can be found deterministically in the interfacial region of two semi-infinite 2D PCs if the system parameters are perturbed from those for conical dispersion condition at **k**=0 and if the bands are derived from monopole and dipole excitations. The dispersion of the interface states can be predicted by the two surface impedances of the PCs on both sides derived by the scattering theory. The necessity for these interface states to exist can be explained by the "character" of band gaps which is governed by the geometric Zak phase of the bulk bands.

**Acknowledgments**

CTC thanks Prof. Feng Wang and Vic Law for discussions. XQ and XM have equal contributions to this work. This work is supported by Hong Kong RGC grants 600311, HKUST2/CRF/11G and AOE/P-02/12.

**References**

[1] A. H. C. Neto, F. Guinea, N. M. R. Peres, K. S. Novoselov and A. K. Geim, Rev. Mod. Phys. **81**, 109 (2009).

[2] K. S. Novoselov, A. K. Geim, S. V. Morozov, D. Jiang, M. I. Katsnelson, I. V. Grigorieva, S. V. Dubonos, and A. A. Firsov, Nature **438**, 197 (2005).

[3] Y. Zhang, Y. W. Tan, H. L. Stormer, and P. Kim, Nature **438**, 201 (2005).

[4] R. A. Sepkhanov, Y. B. Bazaliy, and C. W. J. Beenakker, Phys. Rev. A **75**, 063813 (2007).

[5] T. Ochiai, and M. Onoda, Phys. Rev. B **80**, 155103 (2009).

[6] S. Raghu and F. D. M. Haldane, Phys. Rev. A **78**, 033834 (2008).

[7] F. D. M. Haldane, S. Raghu, Phys. Rev. Lett. **100**, 013904 (2008).

[8] J. Mei, Y. Wu, C. T. Chan, and Z. Q. Zhang, Phys. Rev. B **86**, 035141 (2012).
[9] K. Sakoda, J. Opt. Soc. Am. B **29**, 2770 (2012).
[10] K. Sakoda, Opt. Express **20**, 25181 (2012).

[11] X. Huang, Y. Lai, Z. H. Hang, H. Zheng, and C. T. Chan, Nature Materials **10**, 582 (2011).

[12] F. Liu, Y. Lai, X. Huang, C. T. Chan, Phys. Rev. B **84**, 224113 (2011).

[13] M. Silveirinha and N. Engheta, Phys. Rev. Lett. **97**, 157403 (2006).

[14] R. Liu, Q. Cheng, T. Hand *et al.*, Phys. Rev. Lett. **100**, 023903 (2008).

[15] B. Edwards, A. Alu, M. E. Young, M. Silveirinha, and N. Engheta, Phys. Rev. Let. **100**,

033903 (2008).

[16] R. W. Ziolkowski, Phys. Rev. E **70**, 046608 (2004).

[17] S. Enoch, G. Tayeb, P. Sabouroux, N. Guerin, and P. Vincent, Phys. Rev. Lett. **89**, 213902 (2002).

[18] A. Alu, M. G. Silveirinha, A. Salandrino, and N. Engheta, Phys. Rev. B **75**, 155410 (2007).

[19] J. Hao, W. Yan, and M. Qiu, Appl. Phys. Lett. **96**, 101109 (2010).

[20] Y. Jin and S. He, Opt. Express **16**, 16587 (2010).

[21] V. C. Nguyen, L. Chen, and K. Halterman, Phys. Rev. Lett. **105**, 233908 (2010).

[22] F. J. Lawrence, L. C. Botten, K. B. Dossou, R. C. McPhedran, and C. M. de Sterke, Phys. Rev. A **82**, 053840 (2010).

[23] F. J. Lawrence, L. C. Botten, K. B. Dossou, C. M. de Sterke, and R. C. McPhedran, Phys. Rev. A **80**, 023826 (2009).

[24] Z. Wang, Y. Chong, J. D. Joannopoulos and Marin Soljacic, Nature **461**, 772 (2009).

[25] M. C. Rechtsman, J. M. Zeuner, Y. Plotnik, Y. Lumer, D. Podolsky, F. Dreisow, S. Nolte, M. Segev, and A. Szameit, Nature **496**, 196 (2013).

[26] A. B. Khanikaev, S. H. Mousavi, W. K. Tse, M. Kargarian, A. H. MacDonald and G. Shvets, Nature Mater. **12**, 233 (2013).

[27] L. Lu, L. Fu, J. D. Joannopoulos and M. Soljacic, Nature Photon. **7**, 294 (2013).

[28] K. Fang, Z. Yu, and S. Fan, Nature Photon. **6**, 782 (2012).

[29] M. Hafezi, E. A. Demler, M. D. Lukin, and J. M. Taylor, Nature Phys. **7**, 907 (2011).

[30] Z. Wang, Y. D. Chong, J. D. Joannopoulos, and M. Soljacic, Phys. Rev. Lett. **100,** 013905 (2008).

[31] Z. Yu, G. Veronis, Z. Wang, and S. Fan, Phys. Rev. Lett. **100**, 023902 (2008).

[32] Y. E. Kraus, Y. Lahini, Z. Ringel, M. Verbin, and O. Zilberberg, Phys. Rev. Lett. **109**, 106402 (2012).

[33] Y. Poo, R. X. Wu, Z. Lin, Y. Yang, and C. T. Chan, Phys. Rev. Lett. **106**, 093903 (2011).

[34] Y. Plotnik, M. C. Rechtsman, D. Song, M. Heinrich, J. M. Zeuner, S. Nolte, Y. Lumer, N. Malkova, J. Xu, A. Szameit, Z. Chen and M. Segev, Nature Mater. **13**, 57-62 (2014).

[35] M. C. Rechtsman, Y. Plotnik, J. M. Zeuner, D. Song, Z. Chen, A. Szameit, and M. Segev, Phys. Rev. Lett. **111**, 103901 (2013).

[36] J. Zak, Phys. Rev. Lett. **62**, 2747 (1989).

[37] M. Xiao, Z. Q. Zhang, C. T. Chan, arXiv:1401.1309.

**Figures**

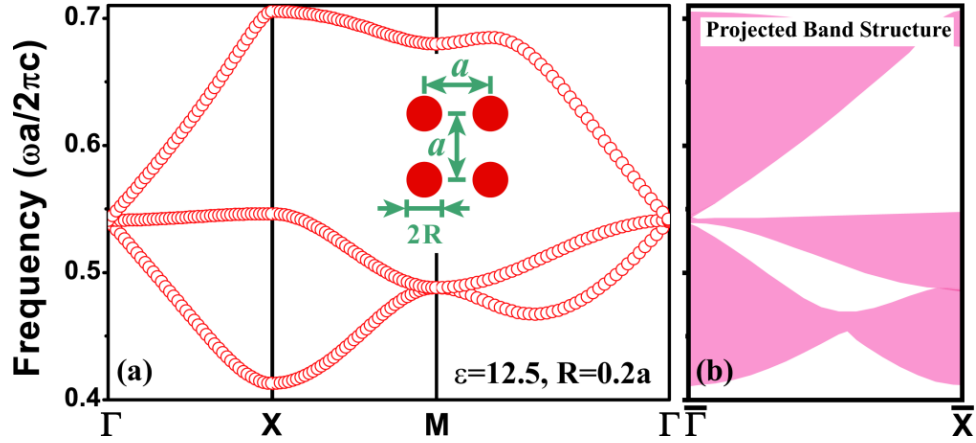

Figure 1 (a) The bulk band structure for $\varepsilon = 12.5$, $R = 0.2a$. The linear bands meeting at **k**=0 has a conical dispersion and a quasi-longitudinal band that is flat along $\Gamma X$. The inset is a schematic picture of a 2D PC consisting an array of dielectric cylinders. (b) The projected band structure for $k_{\parallel}$ along the [10] or [01] direction. Bulk allowed states are shaded in red color. White color marks gaps where surface states can form. The conical dispersion guarantees that there are gaps in the projected band structure above and below the bulk allowed states generated by the flat quasi-longitudinal band.

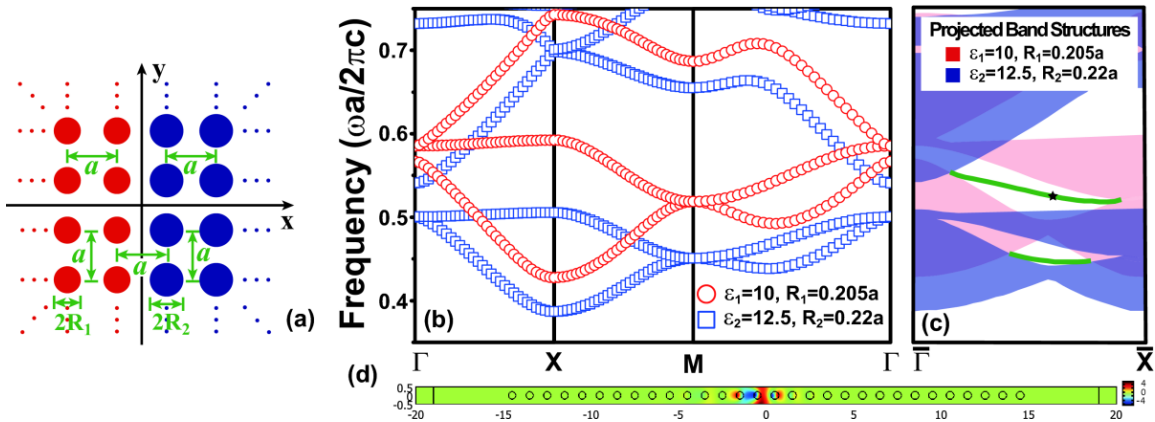

Figure 2 (a) Schematic picture of an interface along the [01] direction constructed by two semi-infinite 2D PCs with a square lattice. The lattice constant for both PCs are *a*. (b) The band structures of 2D PCs with parameters that are close to the Dirac-like cone condition at **k**=0. The red circles are for the PC with $\varepsilon_1 = 10, R_1 = 0.205a$, the blue squares for the PC with $\varepsilon_2 = 12.5, R_2 = 0.22a$. (c) The projected band structures of these two PCs along the interface direction ($\bar{\Gamma}\bar{X}$) with red color for the PC with $\varepsilon_1 = 10, R_1 = 0.205a$ and blue color for the PC with $\varepsilon_2 = 12.5, R_2 = 0.22a$. The green lines in the common partial band gaps represent the interface states. (d) The electric field distribution of the eigen mode of one interface state at the frequency $0.524c/a$ and $k_\parallel = 0.6\pi/a$ (labeled by a black star on the green line in (b)) computed by COMSOL.

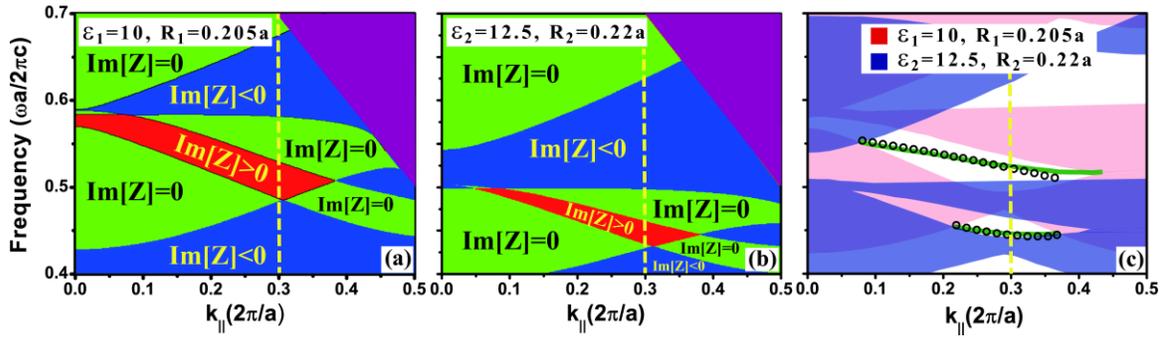

Figure 3 The pass and forbidden band regions in the projected band structures of 2D PCs along the interface direction ($\bar{\Gamma}\bar{X}$) labeled with the imaginary part of the surface impedance ($\text{Im}[Z]$) and the interface states generated by these two PCs. The projected band structures for the PC with (a) $\varepsilon_1 = 10, R_1 = 0.205a$ and (b) $\varepsilon_2 = 12.5, R_2 = 0.22a$. The purple regions represent the regions where higher order diffractions play an important role in scattering theory. (c) The projected band structures and interface states calculated by full-waves calculation. Green lines represent the interface states. Black circles are calculated analytically by surface impedance. The dashed lines are for $k_\parallel = 0.6\pi/a$.

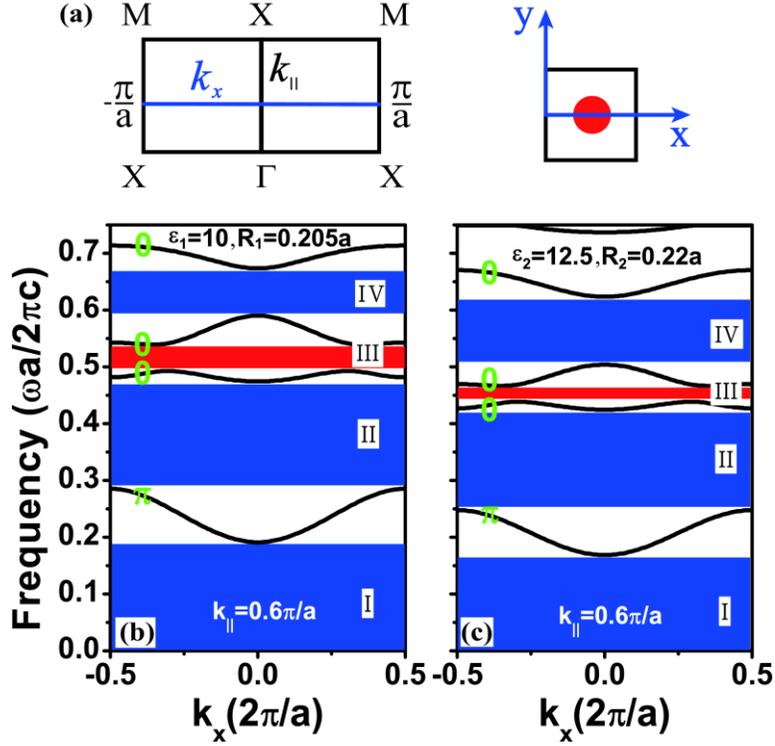

Figure 4 (a) Left panel denotes the first Brillouin zone of a square lattice, the blue line represents $k_x$ varying from $-\pi/a$ to $\pi/a$ with a fixed $k_\parallel$. Since the Zak phase calculation is dependent on the choice of origin, the right panel denotes the coordinate for calculating the Zak phase, the origin is located on the left boundary of the unit cell. The bulk band structure with a fixed $k_y = k_\parallel = 0.6\pi/a$ for the PC with (b) $\varepsilon_1 = 10, R_1 = 0.205a$ and (c) $\varepsilon_2 = 12.5, R_2 = 0.22a$. The Zak phase of the bulk band is either 0 or $\pi$ which is labeled with green color. The blue and red color regions denote different signs of the imaginary part of surface impedance ($\mathrm{Im}[Z]$), blue color for $\mathrm{Im}[Z] < 0$, red color for $\mathrm{Im}[Z] > 0$.

# Appendix

**(I) Interface states in two-dimensional photonic crystals with conical dispersions**

In two-dimensional (2D) photonic crystals (PCs), the Dirac-like cone at **k**=0 can be formed by the accidental degeneracy of the monopole and dipole degrees of freedom [1]. The interface we are considering separates two semi-infinite 2D PCs, each with system parameters (dielectric constant and/or radius of cylinders) that are slightly perturbed from the accidental degeneracy condition to form a Dirac-like cone at the zone center. Due to the requirement of 3 fold degeneracy, we note that the conical dispersion at k=0 (accounting for 2 degrees of freedom) must co-exist with an additional band whose character is quasi-longitudinal and nearly dispersionless near the Dirac-like point. We note that this band is also required to exist because of the zero-index equivalence [1] as zero-index material has an additional longitudinal solution. Here we give two examples in which the frequency difference between two quasi-longitudinal bands is very small as shown in Figs. A1 and A2. Figure A1 shows the case when both PCs on each side have a higher frequency for the dipole mode, whereas in Fig. A2 the monopole modes have a higher frequency. Even though the common partial band gap between two quasi-longitudinal bands is very small for both cases, we are able to find a band of interface states in each common band gap. To visualize these interface states, we also plot in Figs. A1(c) and A2(c) two eigen modes of two interface states at some particular $k_\parallel$ (labeled by black stars on the green lines). It is clearly seen that the electric field is localized near the interfacial region in each case.

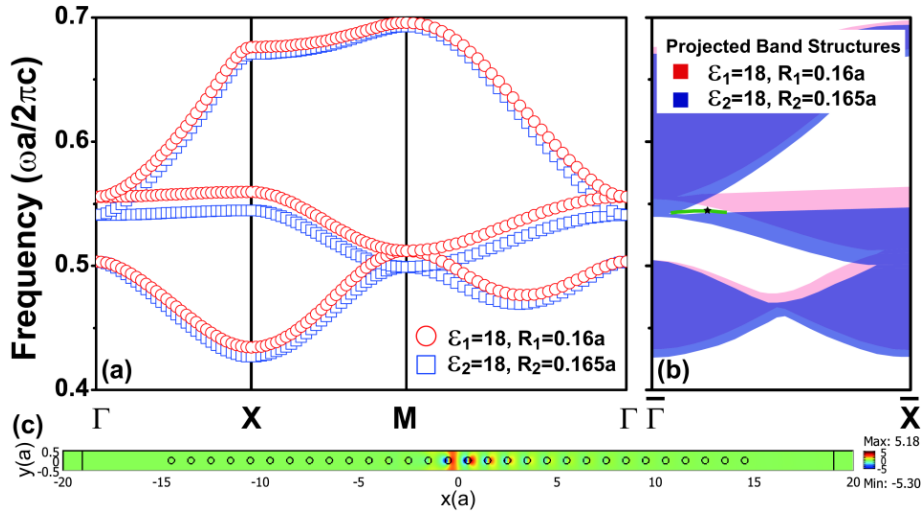

Figure A1. (a) The band structures of 2D PCs with parameters that are close to the conical dispersion condition at **k**=0. The red circles are for the PC with $\varepsilon_1 = 18, R_1 = 0.16a$, the blue squares for the PC with $\varepsilon_2 = 18, R_2 = 0.165a$. (b) The projected band structures of these two PCs along the interface direction ($\overline{\Gamma X}$) with red color for the PC with $\varepsilon_1 = 18, R_1 = 0.16a$ and blue color for the PC with $\varepsilon_2 = 18, R_2 = 0.165a$. The green line in the common partial band gap represents the interface states. (c) The electric field distribution of the eigen mode of one interface state at the frequency $0.544c/a$ and $k_\parallel = 0.22\pi/a$ (labeled by a black star on the green line in (b)) computed by COMSOL.

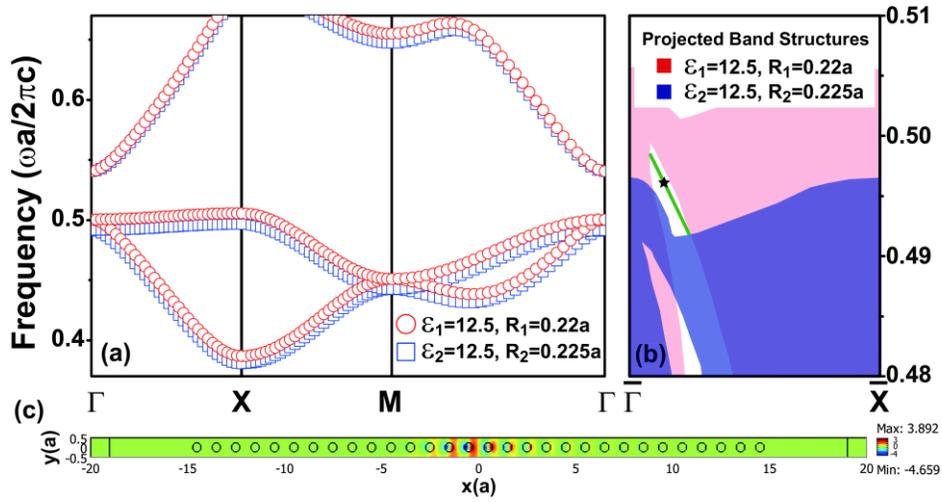

Figure A2. (a) The band structures of 2D PCs with parameters that are close to the conical dispersion condition at **k**=0. The red circles are for the PC with $\varepsilon_1 = 12.5, R_1 = 0.22a$, the blue squares for the PC with $\varepsilon_2 = 12.5, R_2 = 0.225a$. (b) The enlarged projected band structures of these two PCs along the interface direction ($\overline{\Gamma X}$) with red color for the PC with $\varepsilon_1 = 12.5, R_1 = 0.22a$ and blue color for the PC with $\varepsilon_2 = 12.5, R_2 = 0.225a$. The green line in the common partial band gaps represents the interface states. (c) The electric field distribution of the eigen mode of one interface state at the frequency $0.496c/a$ and $k_\parallel = 0.144\pi/a$ (labeled by a black star on the green line in (b)) computed by COMSOL.

From the two examples shown above, we have demonstrated that "perturbing the conical dispersion" generates interface states. Noting that the accidental degeneracy at the conical point at **k**=0 comes from the degeneracy of a pair of doubly degenerate dipole modes and a monopole mode, perturbation will cause the splitting of the 3-fold degeneracy into a 2-fold (dipoles) and 1-fold (monopole) and there are 3 possible combinations:

(i) The monopole is higher in frequency than the dipole at **k**=0 in one PC on one side of the interface and the monopole is lower in frequency than the dipole at **k**=0 on another side of the interface. This case is shown in Fig. 2 of the main text.

(ii) The monopole is lower in frequency than the dipole at **k**=0 in both PCs (See Fig. A3 below). Figure A1 is an extreme case of this category.

(iii) The monopole is higher in frequency than the dipole at **k**=0 in both PCs (See Fig. A4 below). Figure A2 is an extreme case of this category.

In Fig. A3(a), we show the case in which the frequencies of dipole bands are higher than those of monopole bands at **k**=0 on both sides of the interface. The bulk band structures of two PCs have the following parameters: $\varepsilon_1 = 10, R_1 = 0.205a$ for the PC with red circles and $\varepsilon_2 = 18, R_2 = 0.17a$ for the PC with blue squares. The projected band structures of two PCs along the interface direction ($\bar{\Gamma}\bar{X}$) are shown in Fig. A3(b). There are five common partial band gaps: the first one is near the frequency $0.625c/a$, the second one is in the frequency range between $0.526c/a$ and $0.563c/a$, i.e., the region between two quasi-longitudinal flat bands in the bulk band structures of two PCs, the third one is in the range between $0.49c/a$ and $0.51c/a$, the fourth one between $0.46c/a$ and $0.48c/a$ and the fifth one is around the frequency $0.425c/a$. The interface states are found in the second and fourth common partial band gaps denoted by two green lines in Fig. A3(b). The results for the case where the frequencies of monopole bands are higher than those of dipole bands at **k**=0 on either side of the interface are shown in Fig. A4. Interface states are also found.

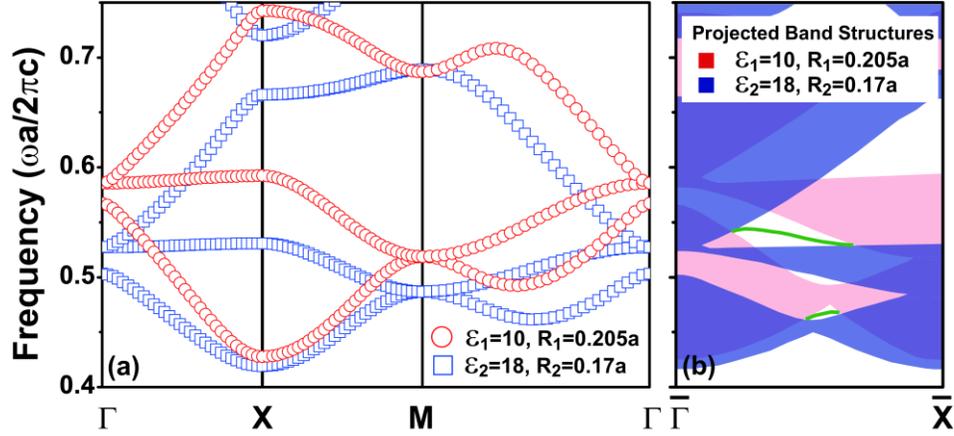

Figure A3. (a) The band structures of 2D PCs with parameters that are close to the conical dispersion condition at **k**=0. The red circles are the band structures for the PC with $\varepsilon_1 = 10, R_1 = 0.205a$, the blue squares for the PC with $\varepsilon_2 = 18, R_2 = 0.17a$. For both of the band structures, the frequencies of dipole bands are higher than those of the monopole bands at **k**=0. (b) The projected band structures of PCs along the interface direction ($\bar{\Gamma}\bar{X}$) with red color for the PC with $\varepsilon_1 = 10, R_1 = 0.205a$ and blue color for the PC with $\varepsilon_2 = 18, R_2 = 0.17a$. There are five common partial band gaps in both of the two projected band structures. The green lines in the common partial band gaps represent the interface states at the interface created by the two PCs. Here, $a$ is the lattice constant of the PCs.

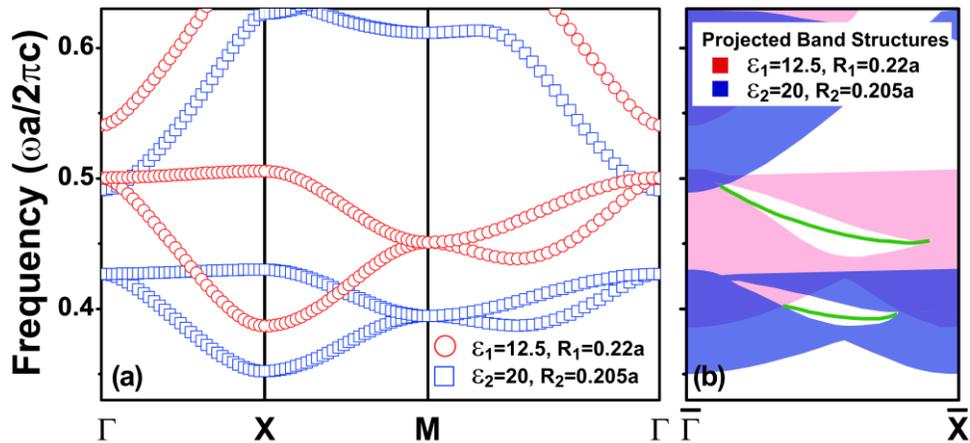

Figure A4. (a) The bulk band structures of two PCs with different parameters that are close to conical dispersion condition at **k**=0. The red circles are the band structures for the PC with $\varepsilon_1 = 12.5, R_1 = 0.22a$, the blue squares for the PC with $\varepsilon_2 = 20, R_2 = 0.205a$. For both of the band structures, the frequencies of dipole bands are lower than those of the monopole bands at **k**=0. (b) The projected band structures of PCs along the interface direction ($\overline{\Gamma X}$) with red color for the PC with $\varepsilon_1 = 12.5, R_1 = 0.22a$ and blue color for the PC with $\varepsilon_2 = 20, R_2 = 0.205a$. There are four common partial band gaps in both of the two projected band structures. The green lines in the common partial band gaps represent the interface states at the interface created by the two PCs. Here, $a$ is the lattice constant of the PCs.

## (II) Surface impedance $Z(\omega, k_\parallel)$ and scattering theory

In order to obtain the surface impedance $Z(\omega, k_\parallel)$ for 2D PC, we use the layer-by-layer scattering formalism, which treats a 2D PC as stacks of 1D PCs. Detailed description of layer-by-layer scattering formalism can be found in the literature [2]. We start with one single constituent layer, which in our system is one single row of cylinders with distance $a$ between the cylinders. Scattering theory allows us to calculate the reflection ($r$) and transmission ($t$) coefficients for this one-layer PC with an incident wave at a given frequency and $k_\parallel$, and in our situation where the scattering comes from the monopole and dipole excitations of the cylinders, the reflection and transmission coefficients have analytic solutions in terms of 1D lattice sums as shown in the main text. The one-layer PC is arranged along $y$-direction with one unit cell along $x$-direction, and the cylinders are centered in the unit cell and at $x = a/2$ (shown in Fig. A5). The $r$ and $t$ for a particular value of $k_\parallel$ determine an impedance for that $k_\parallel$, which we will call $Z_1(\omega, k_\parallel)$ with the subscript "1" denoting the impedance obtained by considering one constituent layer shown in Fig. A5. We can extract the impedance for a N-layer stack by calculating the $r$ and $t$ for a stack of N layers, and let those be denoted by $Z_N(\omega, k_\parallel)$. As long as we choose a centrosymmetry unit cell, and as long as the $r$ and $t$ are determined at the boundary of the unit cell marked by the solid red line, this impedance parameter has a very important property: $Z_N(\omega, k_\parallel)$ is uniquely defined as long as high-order diffraction can be ignored and $Z_N(\omega, k_\parallel)$ has the same value for any number of N from N=1 to ∞ [although $r$ and $t$ depends on thickness] which will simply be denoted as $Z(\omega, k_\parallel)$. More importantly, as $Z(\omega, k_\parallel) = Z_N(\omega, k_\parallel)$ as $N \to \infty$, $Z(\omega, k_\parallel)$ is the surface impedance of the semi-infinite PC for that specific orientation as determined by the layer-stacking.

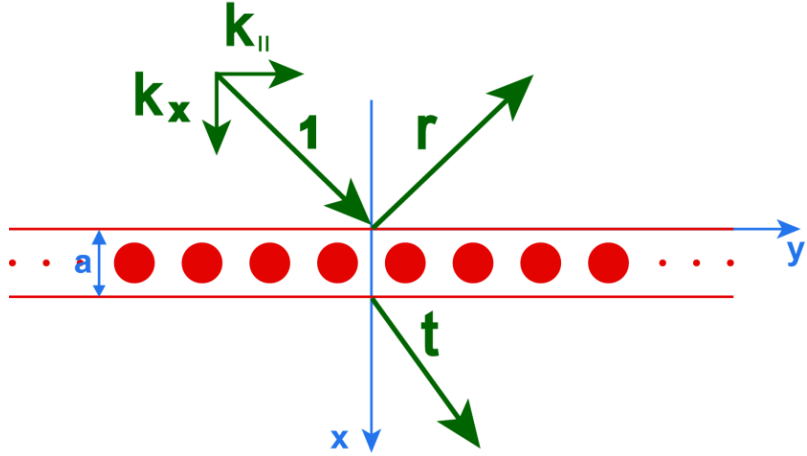

Figure A5. The configuration for calculating the reflection (*r*) and transmission (*t*) coefficients of one-layer PC. The cylinders are at the center of unite cell, arranged along *y*-direction, and the center of cylinders are at $x=a/2$.

We note the surface impedance $Z(\omega,k_{\parallel})$ is extracted via calculating the reflection (*r*) and transmission (*t*) coefficients of the one-layer PC and as such, it ignores evancesent waves. This is a good approximation if high-order inter-layer diffractions can be ignored. To verify whether only considering the zero-order inter-layer diffraction is adequate, we calcualte the projected band structure of one PC along the interface direction ($\overline{\Gamma}\overline{X}$) and compare the result of the full wave calculation with that if high-order inter-layer diffraction can be ignored. The relative permittivity, permeability and radius of the cylinders of the PC are $\varepsilon_1=10, R_1=0.205a$, respectively. The full-wave calculation is shown in Fig. A6(a) and the calculation obtained by scattering theory for only zero-order inter-layer diffraction is shown in Fig. A6(b). The purple region shown in Fig. A6(b) is the region where the higher-order diffractions should play an important role and we see that the results of these two methods agree well with each other for a large region of $k_{\parallel}$.

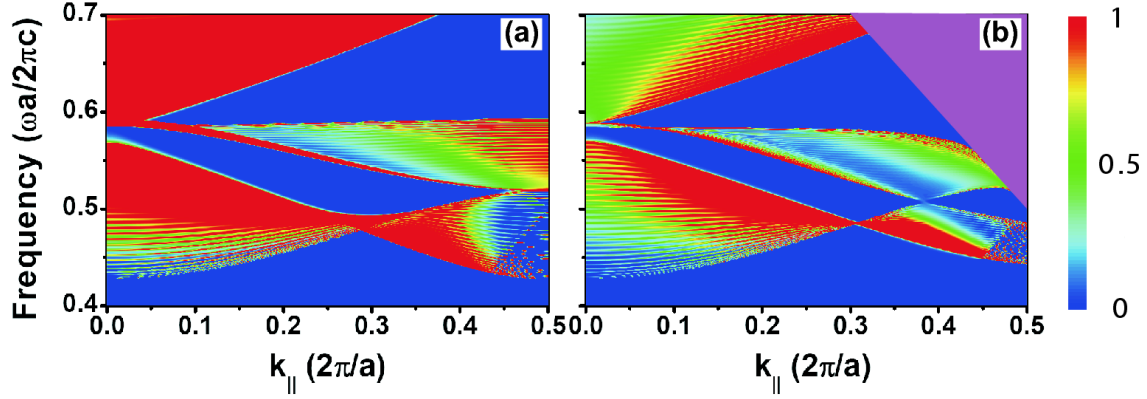

Figure A6. The projected band structures of PC along the interface direction ($\bar{\Gamma}\bar{X}$) with $\varepsilon_1 = 10, R_1 = 0.205a$ calculated by two different methods. (a) for full-wave calculation, (b) for multiple scattering theory with only zero-order inter-layer diffraction. The purple region in (b) represents the higher-order diffractions play an important role in the multiple scattering theory.

## (III) Calculations of the Zak phases.

In the main text, we use the numerical calculations to evaluate the Zak phases of certain reduced 1D bands for a particular $k_{\parallel}$. In this section, we show the details about how to calculate the Zak phase and use the method proposed by W. Kohn [3] to verify the Zak phase obtained by numerical calculations.

In 2D PC, the Bloch wave vector comprises $k_x$ and $k_y$, in order to get the projected band structure, we fix $k_y = k_{\parallel}$ and let $k_x$ vary from $-\pi/a$ to $\pi/a$. In the main text, we have plotted the bulk band structures of two PCs along $k_x$ direction for $k_y = 0.6\pi/a$ (shown in Fig. 4). The definition of Zak phase in 1D is $\varphi_n = i\int_{-\pi/a}^{\pi/a} \left\langle u_{nk_x, k_y=k_{\parallel}} \left| \varepsilon(\vec{r})\partial_{k_x} \right| u_{nk_x, k_y=k_{\parallel}} \right\rangle dk_x$ [4], where $u_{n\vec{k}}$ is the cell periodic part of the Bloch function of the $E$ field for the $n^{th}$ band at a particular $\vec{k}$, $\varepsilon(\vec{r})$ is position dependent relative permittivity in the unit cell. For the implement of numerical calculations, the integral formula has been changed to summation form, so $\varphi_n = -\sum_{l=1}^{M} \text{Im} \ln \left\langle u_{nk_{x,l}, k_y=k_{\parallel}} \left| \varepsilon(\vec{r}) \right| u_{nk_{x,l+1}, k_y=k_{\parallel}} \right\rangle$ in the limit of large M [5]. For TM polarization ($E$ along cylinder axis direction), the eigen modes of $E_{n\vec{k}}$ and $u_{n\vec{k}}$ are related by $E_{n\vec{k}}(\vec{r}) = u_{n\vec{k}}(\vec{r})e^{i\vec{k}\cdot\vec{r}}$. To obtain the Zak phases of these bands, we firstly use COMSOL to calculate the eigen modes $u_{n\vec{k}}(\vec{r})$ of 2D PC for different $k_x$ at a particular $n^{th}$ band and $k_y = k_{\parallel}$. Then, we can calculate the inner product: $\langle u_{nk_{x,I},k_y=k_{\parallel}} | \varepsilon(\vec{r}) | u_{nk_{x,I+1},k_y=k_{\parallel}} \rangle = \iint_{unit\ cell} \varepsilon(\vec{r}) u^*_{nk_{x,I},k_y=k_{\parallel}}(\vec{r}) \cdot u_{nk_{x,I+1},k_y=k_{\parallel}}(\vec{r}) d\vec{r}$. With this inner product, using the periodic gauge $u_{n,-\pi/a,k_y}(x,y) = e^{i2\pi x/a} u_{n,\pi/a,k_y}(x,y)$, the Zak phase of $n^{th}$ band can be obtained (shown in Fig. 4). Since the Zak phase is dependent on the choice of the origin, in the calculation, we set the origin at the left boundary of the unit cell (shown in Fig. 4).

In order to check the numerical results, we will use the method given by W. Kohn [3] to determine the Zak phases of the bulk bands. Noting that our system possesses inversion symmetry and we have effectively an 1D problem after fixing a $k_y = k_{\parallel}$, the Zak phase of the band should be $\pi$ if the eigen modes at the two high symmetry points in the Brillouin zone have different symmetries [3], and it should be zero otherwise. These two high symmetry points in the reciprocal space are the P ($k_x = 0, k_y = k_{\parallel}$) and Q ($k_x = -\pi/a, k_y = k_{\parallel}$) points shown in Fig. A7(a). We choose the origin of the coordinate to be at the center of the cylinder, the unit cell chosen in this way is different from the case depicted Fig. 4 in the main text. We plot the eigen modes ($E_z$ field) of P and Q points in Fig. A7. The real part and imaginary part of the eigen modes represent the interactions between monopole and dipole excitations. Since the system at P and Q points possess $\sigma_x$ mirror symmetry, the real part and imaginary part of the eigen modes should be either even or odd function of $x$. Through analyzing the mirror symmetry of the eigen modes at P and Q points, we can obtain the Zak phases of the bands. Let us first examine the eigen modes of the quasi-longitudinal band (the third band in the band structure shown in Fig. 4) at the P point. The real part and imaginary part of the eigen mode are anti-symmetric (shown in Figs. A7(b) and A7(c)). At the Q point, the real part and imaginary part of the eigen mode are symmetric (shown in Figs. A7(d) and A7(e)). Based on the criterion given by Kohn, the Zak phase for the quasi-longitudinal band is $\pi$. The same analysis shows that the Zak phase for the lower band (the second band in the band structure shown in Fig. 4) is also equal to $\pi$. The Zak phase is dependent on the choice of the origin. If we choose the origin of the coordinate in the middle of two nearest neighbor cylinders (as shown in Fig. 4 in the main text), all the Zak phases of the bands discussed above should have an addition phase of $\pi$. Therefore, both the Zak phases of the second and third bands discussed in the main text should be 0. Through the compatibility relations in group theory with $C_{4v}$ symmetry, we can also analyze the symmetry of the electric fields for the quasi-longitudinal and lower bands at these two high symmetry points [6].

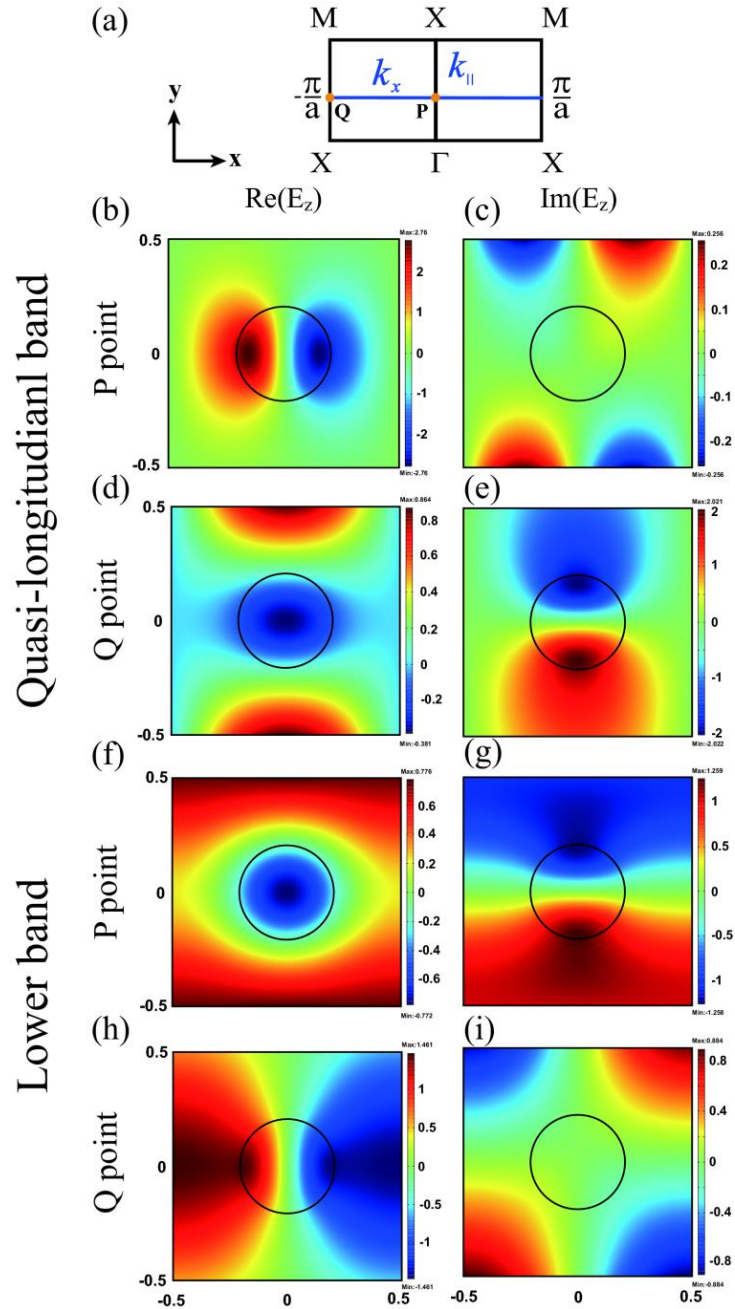

Figure A7. The electric field distributions of the eigen modes at two high symmetry points for two bands in a unit cell. (a) The schematic of $k_x$ from $-\pi/a$ to $\pi/a$ with a fixed $k_\parallel$ in the Brillouin zone. The relative permittivity, permeability and radius of the cylinders of the PC are $\varepsilon_1 = 10, R_1 = 0.205a$, respectively. Here, $a$ is the lattice constant. **P** and **Q** points denote two high symmetry points in the Brillouin zone. (b), (c) and (d), (e) are the electric field distributions of the eigen modes in **P** and **Q** points for the quasi-longitudinal band (the third band in the band structure shown in Fig. 4), respectively. (f), (g) and (h), (i) are the electric field distributions of the eigen modes in **P** and **Q** points for the lower band (the second band in the band structure shown in Fig. 4), respectively. The origin of the coordinate is located at the center of the cylinder.

**(IV) The relationship of the surface impedance and the Zak phase with the interface states**

In the main text and this Appendix, we have given several examples showing the interface states in two semi-infinite 2D PCs. Using scattering theory, the surface impedance $Z(\omega, k_{\parallel})$ of the semi-infinite PC can be obtained. The signs of $\text{Im}(Z(\omega, k_{\parallel}))$ in the gaps of the projected band structure above and below the quasi-longitudinal band are always opposite. For a specific $k_{\parallel}$, in the $\text{Im}(Z(\omega, k_{\parallel})) < 0$ region, the value of $\text{Im}(Z(\omega, k_{\parallel}))$ decreases monotonically from 0 to $-\infty$ with increasing frequency, while in the $\text{Im}(Z(\omega, k_{\parallel})) > 0$ region, the value of $\text{Im}(Z(\omega, k_{\parallel}))$ decreases monotonically from $+\infty$ to 0 with increasing frequency. This implies that $\text{Im}(Z(\omega, k_{\parallel}))$ in the common band gaps between two quasi-longitudinal bands of two PCs can always satisfy the surface wave condition, and hence the existence of one and only one interface state shown in Figs. A1-A4 can be explained. For other common band gaps, as long as they satisfy the $\text{Im}(Z(\omega, k_{\parallel})) < 0$ and $\text{Im}(Z(\omega, k_{\parallel})) > 0$ on either side of the interface, the interface wave existence condition can also be satisfied. Therefore, the interface states must exist (as shown in Figs. A3 and A4). If we know the Zak phase of the bulk band, we do not need to go through the tedious calculation of the scattering problem to obtain $Z(\omega, k_{\parallel})$. Using the band structure information calculated by one unit cell, we can get the Zak phases of the bulk bands, which can determine the characters of the band gaps, and then the existence of the interface states can also be determined. The Zak phase links the bulk band properties to the scattering theory which determines surface properties.

**References**

[1] X. Huang, Y. Lai, Z. H. Hang, H. Zheng, and C. T. Chan, Nature Mater. **10**, 582 (2011).

[2] F. J. Lawrence, L. C. Botten, K. B. Dossou, C. M. Sterke, and R. C. McPhedran, Phys. Rev. A **80**, 023826 (2009).

[3] W. Kohn, Phys. Rev. **115**, 809 (1959).

[4] J. Zak, Phys. Rev. Lett. **62**, 2747 (1989).

[5] R. Resta, J. Phys.: Condens. Matter **12**, R107 (2000).

[6] K. Sakoda, *Optical Properties of Photonic Crystals* 2nd edn (Springer-Verlag, Berlin, Germany, 2004).